# Sub-MHz Self-Q-switching in Nd:LuAG Laser

**Guangju Zhang, Xing Fu, Yijie Shen, and Mali Gong**[*]

*State Key Laboratory of Precision Measurement Technology and Instruments, Department of Precision Instruments, Tsinghua University, Beijing 100084, China*
*\*Corresponding author: gongml@mail.tsinghua.edu.cn*

A compact pulsed Nd:LuAG laser at 1064 nm based on the self-Q-switching technique is reported, having the output power as high as 6.61 W at the incident pump power of 21.32 W, corresponding to the optical conversion efficiency of ~31 %. The temporal width of the pulse was in the range from 532.2 ns to 652.6 ns, and the repetition rate varied between 488.6 kHz and 551.9 kHz. To the best of our knowledge, this is the first report on the self-Q-switching Nd:LuAG laser. Possible reasons for the self-Q-switching of Nd:LuAG were also provided, and the factor leading to the high repetition rate was analyzed. The compact cavity generating pulses with high output power and high repetition rate not only reveal that the self-Q-switching technique could be an efficient method for the generation of pulses with high output power and high repetition rate, but enrich the characteristics of Nd:LuAG crystal.

## 1. INTRODUCTION

High-power, high-repetition-rate pulsed lasers are attractive sources in many applications including remote sensing, bio-medicine, material processing, ranging and so on. Compared with the mode locking technique, Q-switching is able to achieve large pulse energy through a simple cavity-design, thus intense research on the Q-switching technique has been developed in the past decades. Normally, to achieve the Q-switching operation, there is always a modulator in the laser cavity. In the active Q-switching technique, the acoustic-optical or electro-optical modulation modules driven by the periodic external trigger, switch the continuous-wave laser (CW) into pulses [1,2]. However, the active modulator has the drawbacks of high-cost, complexity and inconvenience. As to the passive Q-switching, the saturable absorber (SA) inserted into the cavity is the origin of pulse generation. Therefore SA's performance would significantly influence the operation of the pulsed laser while the cavity can be more compact. However, the complicated fabrication process of the SAs such as SESAMs (semiconductor saturable absorber mirror) still results in high cost [3,4]. Even though a series of new materials have excellent properties of saturable absorption, such as graphene [5-7], carbon nanotube [8], topological insulators [9-11], transition metal dichalcogenides [12-14], black phosphorus [15-17], and gold nanorods [18], their practical characteristics need more time to verify.

In addition to the previous techniques, self-Q-switching (SQS) with no modulators inserted into the cavity, creates bright prospects serving as an effective method for the pulse generation. This technique relies on the gain medium itself to modulate the Q-factor of laser, being free of intracavity loss brought by the inserted modulator, which is beneficial to producing higher average power. On the other hand, the cavity of SQS can be much more compact than the others with much lower costs as well. Therefore the research on the SQS is of great importance and several results have been reported on solid state lasers using SQS. In 2005, Su et al. reported their SQS laser based on the Yb$^{3+}$,Na$^+$:CaF$_2$ single crystal and obtained 1.5 μs pulses at 1050 nm with the repetition rate of 28 kHz and the average power of 400 mW [19]. Gupta et al. demonstrated 460 ns pulses at 914 nm from a SQS Nd:YVO$_4$ laser in 2012, and the corresponding repetition rate was 61.2 kHz with the average output power of 600 mW [20]. In 2014, Xu et al. achieved SQS operation from a Yb: CGB laser in which pulse width of 287 ns at 35 kHz were generated with the two central wavelengths of 1052.6 nm and 1057.7 nm, while the highest output power was 416 mW [21]. Another dual-wavelength SQS laser was reported by Song et al. using a Nd: GYSGG crystal, having the pulse width of 2.02 μs, the repetition rate of 50.2 kHz, and the operating central wavelengths of 1056.86 nm and 1060.23 nm [22]. Liu et al. reported their SQS pulses from the Yb: KGd(WO$_4$)$_2$ system at ~1044 nm. The pulse width and repetition rate were 2.5 μs and 125 kHz respectively, and the average output power was 434.4 mW [23]. In the same year, Cai et al. demonstrated the 1988 nm output with the SQS method from a Tm: YAP laser. The shortest pulse duration was 1.64 μs with the repetition rate of 65.16 kHz and the average output power of 1.68 W [24]. Besides, Zhang et. al. obtained the 1.91 μm pulses from the Tm: YLF laser applying SQS technique in 2018. The corresponding pulse width and repetition rate were 1.4 μs and 21 kHz respectively, and the average output power was 610 mW [25].

It can be concluded from the results above that the SQS technique could be regarded as a promising method in the pulse generation. However, the obtained average powers as well as the repetition rates remained too low to fulfill the application. In the meanwhile, the SQS phenomenon could reveal a new feature of crystals, which is very meaningful and interesting for finding new crystals with same characteristic. Nd:LuAG is a recently grown crystal with several outstanding characteristics. Its long fluorescence time, high thermal conductivity and high saturation fluence make it very effective in the generation of high-energy pulse. In this paper, we demonstrate a pulsed, 1064-nm Nd:LuAG laser applying SQS method for the first time, to the best of our knowledge. The generated pulses have the repetition rate between 488.6 kHz and 551.9 kHz with the temporal duration in the range from 532.2 ns to 652.6 ns. The highest output power was as high as 6.61 W at the incident pump power of 21.32 W, corresponding to the optical conversion efficiency of ~31%. The compact cavity combined with the high repetition rate and the high output power, not only enrich the characteristics of Nd:LuAG crystal, but also reveal the prospect of the SQS technique as the method for producing high-power, high-repetition-rate pulses.

## 2. EXPERIMENTAL SETUP

Figure 1 shows the schematic configuration of the SQS laser, and a compact linear resonator was adopted. The pump source was fiber-coupled laser diode operating at 808 nm. The numerical aperture and the core diameter of the coupled fiber were 0.22 and 105 μm respectively. After travelling through the 1:1 coupling system, the pump light was focused into the gain medium. The gain medium was a Nd:LuAG crystal having the dimensions of 4 mm×10 mm×16 mm and the doping concentration of 0.6 at %. Both surfaces of the crystal were anti-reflection (AR) coated at 808 nm and 1064 nm. The crystal was firstly wrapped with indium foil and then mounted on a cooling system, which consisted of a copper block with the temperature maintained at 20 °C by the deionized water. The input mirror M1 was a plano-concave mirror having the curvature radius of -206.5 mm, and was AR coated at 808 nm and high-reflection (HR) at 1064 nm. M2 was a flat mirror serving as the output coupler with the transmission of 1.4% at 1064 nm. L1 was 15 mm and L2 was 30 mm thus the optical length of the cavity was ~74 mm. M3 was a 45° placed dichroic mirror in order to filter out the residual pump light, while M4 was the beam splitter. To record the temporal characteristics of the output laser, a 1 GHz digital oscilloscope (Tektronix DPO7104C) was used connecting with a fast photodiode detector (Thorlabs DET08C). An optical spectrum analyzer with the resolution of 0.06 nm (Agilent 86140B) was used to obtain the laser spectrum. An OPHIR StarLite power meter [OPHIR 50(150)A-BB-26-QUAD] was utilized to measure the average output power of the laser.

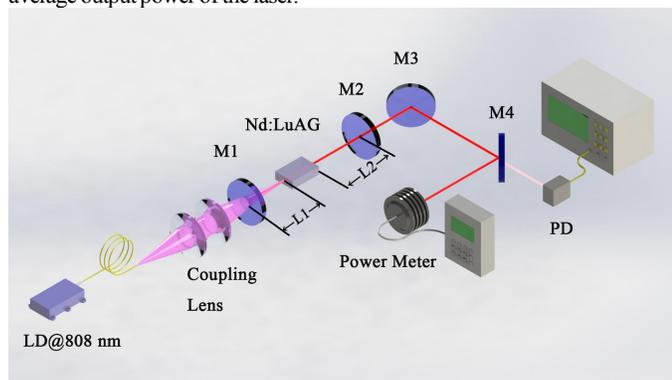

Fig. 1. Experimental setup of the self-Q-switching Nd: LuAG laser. LD: laser diode, PD: photodiode detector.

## 3. RESULTS AND DISCUSSION

Figure 2 shows the average output power of the SQS operation with respect to the incident pump power ranging from 19.51 to 21.32 W, and the average power was 6.61 W at the pump power of 21.32 W, corresponding to the optical conversion efficiency of ~31%. As almost 95 % of the pump power was absorbed by the Nd:LuAG crystal, the detected power by the power meter should be of high precision after the residual pump light was filtered out. Figure 3 shows the evolutions of the pulse width (full width at half maximum, FWHM) and the repetition rate as the incident pump power. With the pump power increasing, the repetition rate increased from 488.6 kHz to 551.9 kHz which was similar to the phenomenon of passive Q-switching. However, the pulse width did not decrease monotonously as the increasing pump power, which might be caused by the thermal perturbation. The shortest pulse width was 532.2 ns with the repetition rate of 521.9 kHz. Figure 4 shows the pulse train at the pump power of 21.32 W, and the corresponding pulse width is 594.3 ns. The pulse energy was 11.98 μJ and the corresponding peak-power was 20.15 W. The optical spectrum of the SQS operation at the pump power of 21.32 W is shown in Figure 5 in which the central wavelength was located at 1064.65 nm. It should be noted that, unlike the previous researches [21,22], the spectrum centered at the single location. By the way, during our experiment, an output coupler with the transmission of 6% was also adopted to observe the SQS operation. However, compared with the 1.4% coupler, the output power was decreased to ~2 W, and the pulse train became unstable.

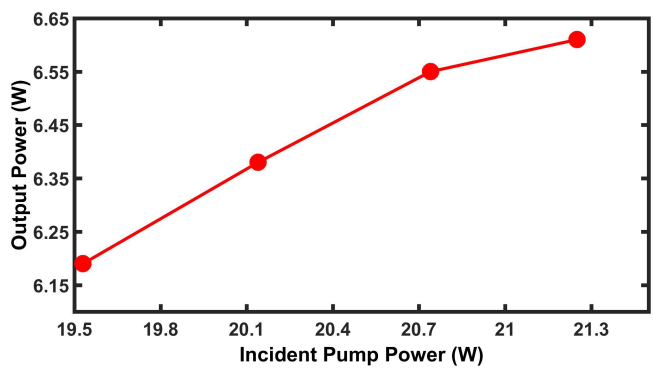

Fig.2 Output power versus the incident pump power.

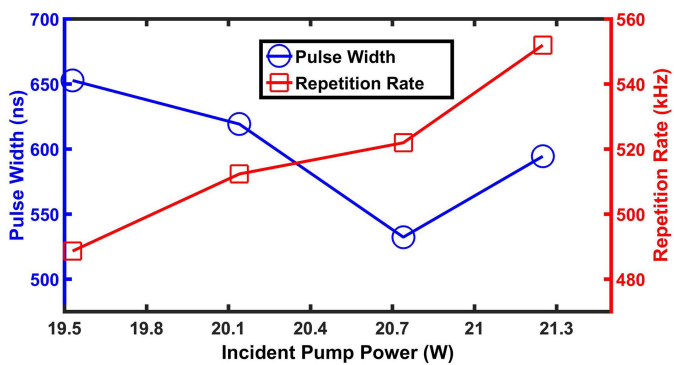

Fig.3 Pulse width and repetition rate as the function of incident pump power.

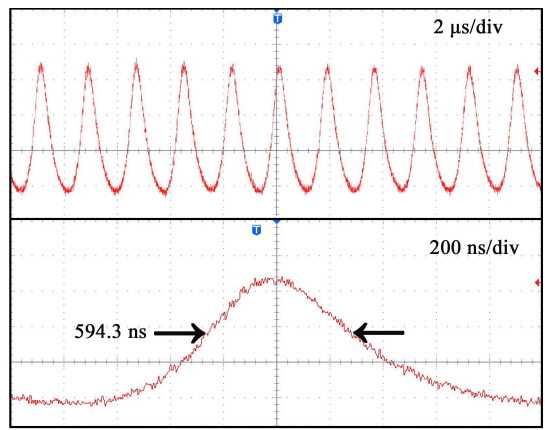

Fig.4 Pulse train and the pulse profile of the self-Q-switching Nd:LuAG laser at the incident pump power of 21.32W.

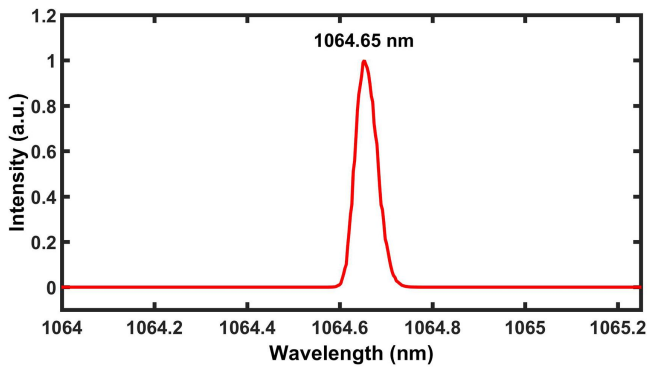

Fig.5 Spectrum of the self-Q-switching Nd: LuAG laser.

Different from the passive Q-switching operation reported in [26, 27], the SQS operation in our experiment was not caused by the saturable absorption of the doping ions such as $Cr^{4+}$. This can be demonstrated by our previous research that with the same crystal the Q-switching pulse could only be obtained after the saturable absorber was inserted into the cavity [28]. For some solid state lasers, the self-Raman effect can generate modulation which could cause SQS operation [29, 30]. During our experiment, we did not observe any other wavelengths but 1064 nm in the spectrum, and this indicates the self-Raman modulation was not the reason. In [20], the reason for the SQS operation was determined as the re-absorption loss in $Nd:YVO_4$ at 914 nm. With the increasing of the intracavity intensity, the re-absorption loss decreases, thus demonstrating an saturable nature for the crystal. Obviously, there should not be strong re-absorption for the Nd: LuAG crystal at 1064 nm, then the re-absorption can not be the reason for the SQS operation.

We believe the SQS operation in the Nd:LuAG laser is attributed to the Kerr lens effect that results from the refractive index changes by the intracavity intensity. As expressed in [31, 32], the refractive index distributes according to the intensity presented as following:

$$n(I) = n_0 + n_2 I \qquad (1)$$

where $n_0$ is the linear refractive index, $n_2$ is the nonlinear refractive index, and $I$ is the power intensity. With this refractive index distribution, when travelling through the Kerr-crystal the light owning higher power will be focused into the smaller size than the lower-power light. Therefore, for a pulse, the Kerr-crystal can be regarded as a lens owning the time-dependent focal length given by following [33]:

$$\frac{1}{f_{NL}(t)} = \frac{4n_2 l P(t)}{\pi \omega^4} \qquad (2)$$

where $l$ is the length of the crystal, $P(t)$ is the time-dependent power during the pulse, and $\omega$ is the effective mode radius on the crystal. With this, the mode radius after being focused by the crystal is also time-dependent during the pulse, and the pulse can be shortened after a soft or hard aperture is inserted into the cavity whose radius is designed according to the focused mode size. This can be described by combining the ABCD method with the focal length of the Kerr lens, and the related nonlinear transmission matrix of the Kerr lens can be expressed as:

$$M_K(t) = \sqrt{1-\gamma(t)} \begin{vmatrix} 1 & \frac{l}{n_0} \\ \frac{-n_0 \gamma(t)}{(1-\gamma(t))l} & 1 \end{vmatrix} \qquad (3)$$

$$\gamma(t) = \left[1 + \frac{1}{4}\left(\frac{2\pi n_0 \omega^2}{\lambda l} - \frac{\lambda l}{2\pi n_0 \omega_0^2}\right)^2\right]^{-1} \frac{P(t)}{P_c} \qquad (4)$$

$$P_c = \frac{c\varepsilon_0 \lambda^2}{2\pi n_2} \qquad (5)$$

where, $\lambda$ is the laser wavelength, $l$ is the length of the Kerr crystal, $\omega_0$ is the beam waist radius calculated at $P(t)/P_c = 0$, $P_c$ is the critical power for Kerr lens effect, $\varepsilon_0$ is the dielectric permeability of vacuum, $n_2$ is the nonlinear refractive index of the Kerr crystal, $c$ is the speed of light. The effective mode radius on the crystal can be obtained according to the ABCD propagation laws [34]:

$$\omega(t)^2 = \frac{2\lambda B(t)}{\pi \sqrt{4-[A(t)+D(t)]^2}} \qquad (6)$$

where $A(t)$, $B(t)$ and $D(t)$ are the propagation matrix elements which are time-dependent and directly affected by $P(t)$. From Equation (6) we can see there will exist the temporal variation of the effective spot size compared to the fixed radius of the hard or soft aperture thus providing the modulation in the laser. Then the aperture can be regarded as a fast saturable absorber which will passively Q-switch the laser system. In our laser, the mode radius located at the pump waist was about 190 μm and the ratio between the mode radius and the pump waist was ~3.6 which indicates that the pump waist could act as a soft aperture in our laser. As given in [35,36], the pulse width and the repetition rate of the passive Q-switched pulse can be expressed as following:

$$\tau_p \approx \frac{3.52 T_R}{M_0} \qquad (7)$$

$$f_{rep} = \frac{1}{\tau} \cdot \frac{G_0}{M_0} \qquad (8)$$

where, $\tau_p$ is the pulse width, $T_R$ is the cavity round-trip time, $M_0$ is the small signal loss brought by the modulator, $f_{rep}$ is the repetition, $\tau$ is the fluorescence time, and $G_0$ presents the small signal gain of the trip around the cavity. It can be seen that both the pulse width and the repetition rate are affected by the cavity length and the intracavity modulation caused by the soft aperture. Therefore, to achieve shorter pulses, the cavity length should be shorter while the modulation depth should be larger. On the other hand, although the Nd:LuAG crystal has a long fluorescence time of 277 μs, as there were no saturable absorbers, very low loss was introduced into the cavity, which could bring the system the high repetition rate. Further research will focus on the optimization of the laser system to achieve shorter pulses with the higher repetition rate.

## 4. CONCLUSION

In conclusion, a compact self-Q-switching Nd:LuAG laser at 1064 nm owning high output power and high repetition rate was achieved. The highest average output power was 6.61 W at the incident pump power of 21.32 W, and the corresponding optical conversion efficiency was ~31 %. The temporal width of the pulse was in the range from 532.2 ns to 652.6 ns, and the repetition rate varied from 488.6 kHz to 551.9 kHz. To the best of our knowledge, this is the first report on the self-Q-switching Nd:LuAG laser. In the paper, possible reasons for the self-Q-switching of Nd:LuAG was provided, and the factors affecting the pulse width and the repetition rate were also analyzed. The compact cavity generating pulses with high output power and high repetition rate gives the self-Q-switching technique a bright prospect for the pulse generation, and the experimental phenomenon also enriches the characteristics of Nd:LuAG crystal.

## ACKNOWLEDGMENT

This work was supported by the National Key Research and Development Program of China (Grant No. 2017YFB1104500).

## REFERENCES

1. X. Yan, Q. Liu, X. Fu, H. Chen, M. Gong, and D. Wang, "High repetition rate dual-rod acousto-optics Q-switched composite Nd:YVO4 laser," Opt. Express **17**, 21956-21968 (2009).

2. Q. Liu, M. Gong, H. Wu, F. Lu, and C. Li, "Electro-optic Q-switched Yb:YAG slab laser," Laser Phys. Lett. **3**, 249-251 (2006).
3. L. Sun, L. Zhang, H.J. Yu, L. Guo, J.L. Ma, J. Zhang, W. Hou, X.C. Lin, and J.M. Li, "880 nm LD pumped passive mode-locked TEM00 Nd:YVO$_4$ laser based on SESAM," Laser Phys. Lett. **7**, 711-714 (2010).
4. Z. Cong, D. Tang, W. Tan, J. Zhang, C. Xu, D. Luo, X. Xu, D. Li, J. Xu, X. Zhang, and Q. Wang, "Dual-wavelength passively mode-locked Nd:LuYSiO$_5$ laser with SESAM," Opt. Express **19**, 3984-3989 (2011).
5. H. Zhang, D. Y. Tang, L. M. Zhao, Q. L. Bao, and K. P. Loh, "Large energy mode locking of an erbium-doped fiber laser with atomic layer graphene," Opt. Express **17**, 17630-17635 (2009).
6. Y. Song, L. Li, H. Zhang, D. Shen, D. Tang, and K. Loh, "Vector multi-soliton operation and interaction in a graphene mode-locked fiber laser," Opt. Express **21**, 10010-10018 (2013).
7. H. Zhang, D. Tang, L. Zhao, Q. Bao, and K. Loh, "Vector dissipative solitons in graphene mode locked fiber lasers," Opt. Commun. **283**, 3334-3338 (2010).
8. F. Wang, A. G. Rozhin, V. Scardaci, Z. Sun, F. Hennrich, I. H. White, W. I. Milne, and A. C. Ferrari, "Wideband-tuneable, nanotube mode-locked, fibre laser," Nature Nanotechnology **3**, 738-742 (2008).
9. P. Li, G. Zhang, H. Zhang, C. Zhao, J. Chi, Z. Zhao, C. Yang, H. Hu, and Y. Yao, "Q-Switched Mode-Locked Nd: YVO$_4$ Laser by Topological Insulator Bi$_2$Te$_3$ Saturable Absorber." IEEE Photon. Technol. Lett. **26**, 1912-1915 (2014).
10. J. Li, H. Luo, L. Wang, C. Zhao, H. Zhang, H. Li, and Y. Liu, "3μm mid-infrared pulse generation using topological insulator as the saturable absorber," Opt. Lett. **40**, 3659-3662 (2015).
11. M. Liu, N. Zhao, H. Liu, X. Zheng, A. Luo, Z. Luo, W. Xu, C. Zhao, H. Zhang, and S. Wen, "Dual-Wavelength Harmonically Mode-Locked Fiber Laser With Topological Insulator Saturable Absorber," IEEE Photon. Technol. Lett., **26**, 983-986 (2014).
12. D. Mao, Y. Wang, C. Ma, L. Han, B. Jiang, X. Gan, S. Hua, W. Zhang, T. Mei, J. Zhao, " WS$_2$ mode-locked ultrafast fiber laser," Sci. Rep. **5**, 7965 (2015).
13. M. Liu, X. Zheng, Y. Qi, H. Liu, A. Luo, Z. Luo, W. Xu, C. Zhao, and H. Zhang, " Microfiber-based few-layer MoS$_2$ saturable absorber for 2.5 GHz passively harmonic mode-locked fiber laser," Opt. Express **22**, 22841-22846 (2014).
14. Z. Luo, D. Wu, B. Xu, H. Xu, Z. Cai, J. Peng, J. Weng, S. Xu, C. Zhu, F. Wang, Z. Sun, and H. Zhang, " Two-dimensional material-based saturable absorbers: towards compact visible-wavelength all-fiber pulsed lasers," Nanoscale **8**, 1066-1072 (2016).
15. X. Wang, Zhe. Wang, Y. Wang, L. Li, G. Yang and J. Li, "Watt-level high-power passively Q-switched laser based on a black phosphorus solution saturable absorber." Chin. Opt. Lett. **15**, 011402 (2017).
16. H. Mu, S. Lin, Z. Wang, S. Xiao, P. Li, Y. Chen, H. Zhang, H. Bao, S. Lau, C. Pan, D. Fan, and Q. Bao, " Black Phosphorus-Polymer Composites for Pulsed Lasers," Adv. Opt. Mater. **3**, 1447-1453 (2015).
17. Z. Qin, G. Xie, H. Zhang, C. Zhao, P. Yuan, S. Wen, and L. Qian, "Black phosphorus as saturable absorber for the Q-switched Er:ZBLAN fiber laser at 2.8 μm," Opt. Express **23**, 24713-24718 (2015).
18. H. Zhang, and J. Liu, " Gold nanobipyramids as saturable absorbers for passively Q-switched laser generation in the 1.1 μm region." Opt. Lett. **41**, 1150-1153 (2016).
19. L. Su, and J. Xu, " Low-threshold diode-pumped Yb$^{3+}$,Na$^+$:CaF$_2$ self-Q-switched laser," Opt. Express **13**, 5635-5640 (2005).
20. P. K. Gupta, A. Singh, S. K. Sharma, P. K. Mukhopadhyay, K. S. Bindra, and S. M. Oak, " Note: Self Q-switched Nd:YVO$_4$ laser at 914 nm," Rev. Sci. Instrum. **83**, 046110 (2012).
21. J. Xu, Y. Ji, Y. Wang, Z. You, H. Wang, and C. Tu, " Self-Q-switched, orthogonally polarized, dual wavelength laser using long-lifetime Yb$^{3+}$ crystal as both gain medium and saturable absorber," Opt. Express **22**, 6577-6585 (2014).
22. Q. Song, G. Wang, B. Zhang, Q. Zhang, W. Wang, M. Wang, G. Sun, Y. Bo, and Q. Peng, " Dual-wavelength self-Q-switched Nd:GYSGG laser," Mod. Opt. **62**, 353-357 (2014).
23. J. Liu, J. Tian, Z. Dou, M. Hu, and Y. Song, " Self-Q-switching in bulk Yb:KGd(WO$_4$)$_2$ laser," Chin. Opt. Lett. **13**, 061407 (2015).
24. W. Cai, J. Liu, C. Li, H. Zhu, P. Ge, L. Zheng, L. Su, and J. Xu, " Compact self-Q-switched laser near 2 μm," Opt. Commun. **334**, 287-289 (2015).
25. B. Zhang, L. Li, C.J. He, F.J. Tian, X. T. Yang, J. H. Cui, J. Z. Zhang, and W. M. Sun, " Compact self-Q-switched Tm:YLF laser at 1.91 μm," Optics and Laser Technology **100**, 103–108 (2018).
26. J. Dong, P. Deng, Y. Lu, Y. Zhang, Y. Liu, J. Xu, and W. Chen, " Laser-diode-pumped Cr$^{4+}$, Nd$^{3+}$:YAG with self-Q-switched laser output of 1.4 W," Opt. Lett. **25**, 1101-1103 (2000).
27. S. Wang, Q. Li, S. Du, Q. Zhang, Y. Shi, J. Xing, D. Zhang, B. Feng, Z. Zhang, and S. Zhang, " Self-Q-switched and mode-locked Nd,Cr:YAG laser with 6.52-W average output power," Opt. Commun. **277**, 130-133 (2007).
28. G. Zhang, T. Liu, Y. Shen, C. Zhao, B. Huang, Z. Kang, G. Qin, Q. Liu, and X. Fu, " 516 mW, nanosecond Nd:LuAG laser Q-switched by gold nanorods," Chin. Opt. Lett. **16**, 020011 (2018).
29. C. Y. Lee, C. C. Chang, H. C. Liang, and Y. F. Chen, " Frequency comb expansion in a monolithic self-mode-locked laser concurrent with stimulated Raman scattering," Laser Photonics Rev. **8**, 750-755 (2014).
30. R. Frey, A. de Martino, and F. Pradbre, " High-efficiency pulse compression with intracavity Raman oscillators," Opt. Lett. **8**, 437-439 (1983).
31. R. H. Stolen, and A. Ashkin, " Optical Kerr effect in glass waveguide." Appl. Phys. Lett, **22**, 294–296 (1973)
32. M. Sheik-Bahae, D. C. Hutchings, D. J. Hagan, and E. W. V. Stryland, " Dispersion of bound electron nonlinear refraction in solids," IEEE J. Quantum Electron. **27**, 1296–1309 (1991),
33. Shai Yefet, and Avi Pe'er, " A Review of Cavity Design for Kerr Lens Mode-Locked Solid-State Lasers," Appl. Sci. **3**, 694-724 (2013).
34. Z. Li, J. Peng, J. Yao, and M. Han, " The characteristics of Kerr-lens mode-locked self-Raman Nd:YVO$_4$ 1176 nm laser," Opt. Laser Technol. **89,** 1-5 (2017).
35. E. Beyatli, A. Sennaroglu, and U. Demirbas, " Self-Q-switched Cr:LiCAF laser," J. Opt. Soc. Am. B **30**, 914-921 (2013).
36. G. J. Spühler, R. Paschotta, R. Fluck, B. Braun, M. Moser, G. Zhang, E. Gini, and U. Keller, " Experimentally confirmed design guidelines for passively Q-switched microchip lasers using semiconductor saturable absorbers," J. Opt. Soc. Am. B **16**, 376-388 (1999).